%% file: ShortSummary.tex
\documentclass[12pt]{elsarticle}

\RequirePackage{lineno}
\usepackage{subfigure}
\usepackage[percent]{overpic}
\usepackage{color}
\usepackage{units}
\usepackage[footnotesize]{caption}
\usepackage[colorlinks=true,linkcolor=DarkBlue,citecolor=DarkBlue]{hyperref}
\usepackage{fancyhdr}
\usepackage[percent]{overpic}
\usepackage{wasysym}

\IfFileExists{NewCommands.tex}       {\input{NewCommands}}       {}

\makeatletter
\def\ps@pprintTitle{%
 \let\@oddhead\@empty
 \let\@evenhead\@empty
 \def\@oddfoot{}%
 \let\@evenfoot\@oddfoot}
\makeatother

\begin{document}
\begin{frontmatter}

\title{PINGU: \\ A Vision for Neutrino and Particle Physics at the South Pole}

\input{AuthorList}

\begin{abstract}

  The Precision IceCube Next Generation Upgrade (PINGU) is a proposed
  low-energy in-fill extension to the IceCube Neutrino Observatory.
  With detection technology modeled closely on the successful IceCube
  example, PINGU will provide a 6~Mton effective mass for neutrino
  detection with an energy threshold of a few GeV.  With an
  unprecedented sample of over 60,000 atmospheric neutrinos per year
  in this energy range, PINGU will make highly competitive
  measurements of neutrino oscillation parameters in an energy range
  over an order of magnitude higher than long-baseline neutrino beam
  experiments.  PINGU will measure the mixing parameters $\thTT$ and
  $\dmTT$, including the octant of $\thTT$ for a wide range of values,
  and determine the neutrino mass ordering at $3\sigma$ median
  significance within $\YrsToThreeSigmaSystLimited$~years of
  operation.  PINGU's high precision measurement of the rate of
  $\nutau$ appearance will provide essential tests of the unitarity of
  the $3\times 3$ PMNS neutrino mixing matrix.  PINGU will also
  improve the sensitivity of searches for  low
  mass dark matter in the Sun, use neutrino tomography to directly probe the composition of
  the Earth's core, and improve IceCube's sensitivity to neutrinos
  from Galactic supernovae.  Reoptimization of the PINGU design has
  permitted substantial reduction in both cost and logistical
  requirements while delivering performance nearly identical to
  configurations previously studied. 
\end{abstract}

\end{frontmatter}




\section*{Introduction}
Following the discovery of neutrino oscillations which show that
neutrinos have mass~\cite{Fukuda:1998mi,Ahmad:2001an}, experiments using neutrinos
produced in the atmosphere, in the sun, at accelerators, and at
reactors have measured the mixing angles and mass-squared differences
that characterize the oscillations between the three known flavors of
neutrinos.  Several important questions remain: whether the mixing
angle $\thTT$ is maximal and, if not, whether $\thTT < 45^\circ$ or
$\thTT > 45^\circ$ (the ``octant'' of $\thTT$), whether the ordering
of the mass eigenstates is ``normal'' or ``inverted'', and whether
Charge-Parity (CP) symmetry is violated with nonzero $\dcp$ in the
lepton sector.  More fundamentally, a better understanding of neutrino
oscillations may shed light on the origins of neutrino mass, the
possible relationship of neutrinos to the matter-antimatter asymmetry
of the universe, and probe new physics beyond the Standard Model.

The Precision IceCube Next Generation Upgrade (PINGU) will provide
unprecedented sensitivity to a broad range of neutrino oscillation
parameters.  Embedded in the existing IceCube/DeepCore subarray, with
an energy threshold of less than 5~GeV, PINGU will make highly
competitive measurements of atmospheric mixing parameters, the octant
of $\thTT$, $\nutau$ appearance, and the neutrino mass ordering (NMO,
also referred to as the neutrino mass hierarchy), through studies of a
range of neutrino energies and path lengths which cannot be probed by
long-baseline or reactor neutrino experiments.  PINGU will also
improve the sensitivity of IceCube to neutrino bursts from supernovae
and to neutrinos produced by dark matter annihilations.

In the past few years, in addition to the discovery of high energy
neutrinos of astrophysical origin~\cite{Aartsen:2013pza}, the IceCube
Collaboration has made competitive measurements of neutrino
oscillations~\cite{Aartsen:2013jza,Aartsen:2014yll} and searches for
dark matter~\cite{Aartsen:2012kia}.  The technologies for drilling
holes, deploying instruments, and detecting neutrinos in the
deep Antarctic ice are proven, and the  costs and risks
of constructing PINGU are moderate and well understood.   As an
extension of the IceCube detector, the incremental operational costs of
PINGU would be correspondingly low.

\section*{The South Pole Station and the IceCube Neutrino Observatory}

\begin{figure}
    \begin{center}
      \includegraphics[width=12cm, angle=0]{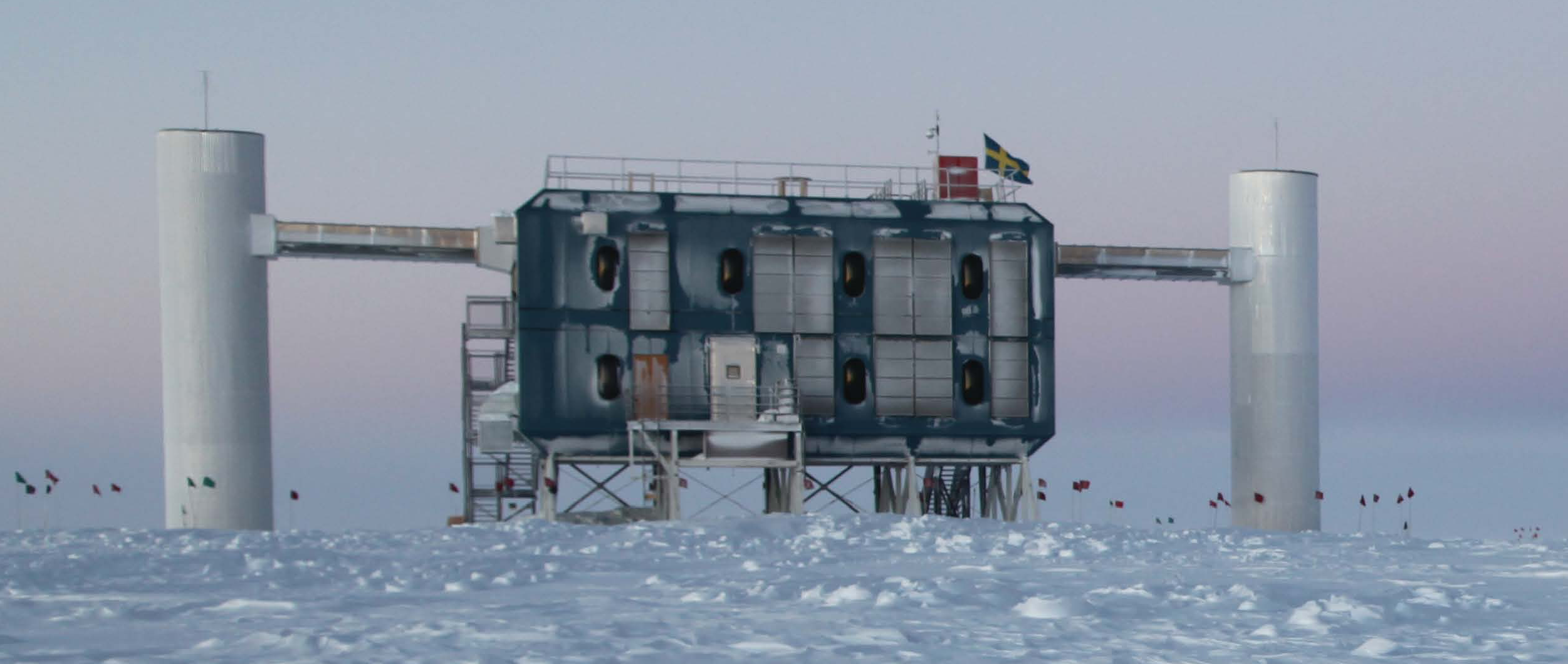}
   \end{center}
    \caption{The IceCube Laboratory building houses power,
      communications and data acquisition systems for IceCube and other
      experiments at the South Pole (photo by S. Lidstr\"om, IceCube/NSF).}
    \label{fig:ICL_ExecSumm}   
 \end{figure}

Over the past decade, the South Pole has emerged as a world-class site
for astronomy, particle astrophysics and neutrino oscillation physics.
At the Amundsen-Scott South Pole Station the glacial ice is more than
2.8~km thick, radiopure, and optically clear~\cite{Aartsen:2013rt},
enabling the construction of a neutrino telescope of unprecedented
scale.  The IceCube Neutrino Observatory, the world's largest neutrino
detector, has been in full operation since 2011.  IceCube uses 5160
optical sensors attached to 86 vertical ``strings'' (cables) to
transform one billion tons of Antarctic ice into a Cherenkov radiation detector.
The sensor modules were deployed using a hot water drill to melt
holes 2.5~km deep in the ice, with the modules deployed at depths of
1.5--2.5~km below the surface.  The NSF's Amundsen-Scott Station
provides comprehensive infrastructure for IceCube's scientific activities,
including the IceCube Laboratory building that houses power,
communications, and data acquisition systems, shown in Fig.~\ref{fig:ICL_ExecSumm}.

The Antarctic ice cap permits very large volumes of material to be
instrumented at relatively low cost.  DeepCore, the low energy
subarray of IceCube, is located at the bottom center of the array and
observes some 20,000 neutrinos per year at energies below 50~GeV,
incident from all directions.  The temperature and radiopurity of the
ice greatly reduce thermionic and radioactive noise in the
photomultiplier tubes (PMTs), the fundamental building block of the
IceCube detector, aiding in the observation of lower energy neutrinos.
The outermost IceCube sensors detect and enable an active veto of
incoming atmospheric muons, reducing muon background rates in the deep
detector to levels comparable to those in deep mines.

\section*{PINGU Design}

PINGU will greatly enhance IceCube's capabilities below a neutrino
energy of 50 GeV with
the deployment of additional photodetector modules within DeepCore,
over an instrumented volume of 6
Mton.  With an energy threshold of a few GeV, PINGU will
substantially improve precision for neutrino events
below 20~GeV --- the key energy range for measurements of the
atmospheric neutrino oscillation patterns and detection of the
imprint of the neutrino mass ordering on these patterns.  PINGU
has a number of attractive features:
\begin{itemize}[\raisebox{0.25ex}{\tiny$\bullet$}]
\item  No state-of-the-art development required
\item  $>10$ years experience of IceCube installation and operations
\item Performance and cost estimates based on existing detector and
  tools
\item Low marginal cost of operations, leveraging IceCube
  infrastructure
\item Near 100\% duty factor
\item $>60,000$ neutrino events/year
\item $>3,000 \; \nu_\tau$ events per year 
\item Broad sensitivity to new  physics through observation of a wide
  range of neutrino energies and baselines
\end{itemize} 


PINGU leverages the experience gained from designing, deploying and
operating IceCube, enabling a rapid construction time with minimal
risk and at relatively modest expense.  The recent development of the
capability to deliver cargo and fuel to the station via overland
traverse rather than aircraft, as well as planned improvements in
drilling efficiency and sensor power requirements, make the logistical
and operational footprint of PINGU significantly smaller than that of
IceCube both during and after construction. 

\begin{figure}
  \begin{center}
    \includegraphics[width=6in]{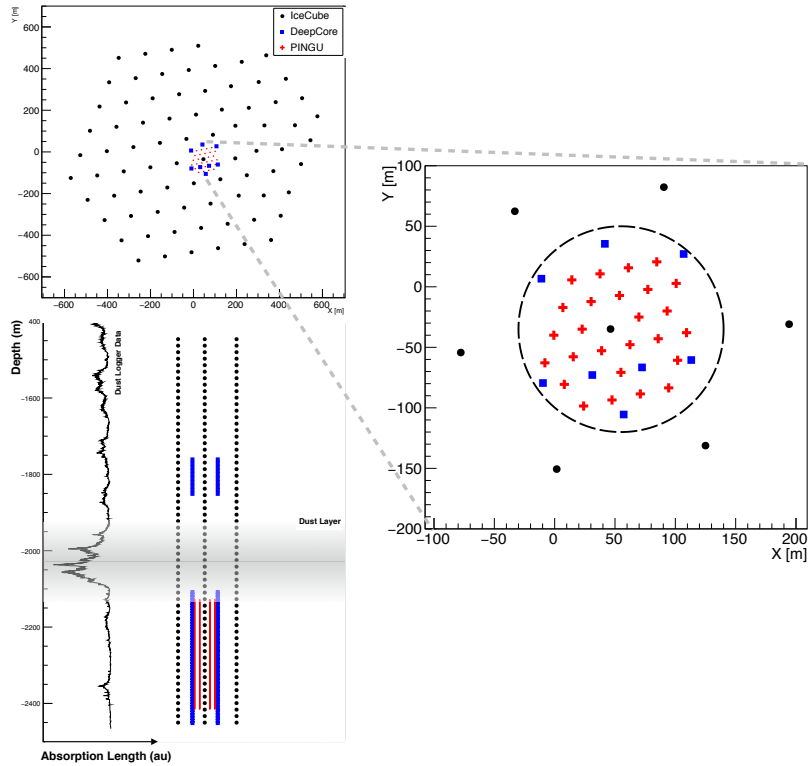}
  \end{center}
  \caption{Schematic layout of PINGU within the IceCube DeepCore
    detector.  In the top view inset at right, black circles mark
    standard IceCube strings, on a \unit[125]{m} hexagonal grid.  Blue
    squares indicate existing DeepCore strings, and red crosses show
    proposed PINGU string locations.  PINGU modules would be deployed
    in the clearest ice at the bottom of the detector, as shown in the
  vertical profile at bottom, with vertical spacing several times
  denser than DeepCore.}
  \label{Fig:schematic_ExecSumm}
\end{figure}

Initial studies of PINGU performance~\cite{Aartsen:2014oha} showed
that PINGU would deliver a world-class 6~Mton water Cherenkov detector
for a cost below US\$100M.  Those
projections were based on a configuration of 40 new strings, each
mounting 96 optical modules.  Our recent studies have shown that a geometry
that concentrates a slightly larger amount of PMT photocathode area
on fewer strings provides the same sensitivity while reducing both
costs and logistical support requirements significantly.  A schematic
of this design is shown in Fig.~\ref{Fig:schematic_ExecSumm}.  Based
on our experience with IceCube, in which 18--20 strings were deployed
per season once construction was underway, 26 strings of 192 optical
modules each could be installed at the South Pole in two deployment
seasons.  This configuration would provide nearly identical
performance to the original 40-string design.  Even a reduced 20
string geometry, which could be deployed in two seasons with 
considerable schedule contingency, would still enable the essential scientific
program, even though it would provide less precise event
reconstruction and reduced performance compared to the
projections presented here.

The studies presented in this document are based on the new, less expensive 26-string configuration. In this configuration, PINGU will be composed of sensors similar in shape and size to those
already deployed in IceCube, enabling deployment with nearly identical
techniques and equipment.  For the purposes of this
study, a sensor identical to the current IceCube
DOM~\cite{Abbasi:2008aa,Abbasi:2010vc} has been assumed.  This would
require only modest  updates to the electronics to be used in PINGU.
We are also evaluating the possibility of replacing the
optical modules with multi-PMT mDOMs~\cite{Classen:2013xma,Adrian-Martinez:2014vja}.
A string consisting of 125 mDOMs
would provide 40\% more photocathode area, as well as directional
information on the arriving photons, for the same cost as a string of
192 regular optical modules. This promises further potential improvements over current performance projections.

The
existing IceCube DOMs that will surround PINGU will provide a highly effective active veto
against downward-going cosmic ray muons, the chief background for all
PINGU physics channels, a strategy successfully developed for DeepCore
measurements~\cite{Aartsen:2014yll}.  The surrounding instrumentation
will also provide containment of muons up to $E_\mu \sim 100$ GeV, improving
energy resolution and utilizing the existing IceCube detector to
substantially improve PINGU's performance relative to a stand-alone
instrument.  PINGU will be designed as an extension of IceCube,
closely integrated with IceCube's online and offline systems, leading
to a very low incremental cost of operation.

PINGU will provide an effective detector target mass of 6~Mton for
$\numu$ charged-current interactions, fully efficient above 8~GeV and
 50\% efficient at $\sim$3~GeV, yielding data samples of approximately
65,000 upgoing neutrinos per year at energies below 80~GeV.  On average, a 10
GeV $\numu$ CC event will produce 90 Cherenkov photons detected by
PINGU; existing IceCube reconstruction algorithms applied to simulated
PINGU events yield an energy resolution $\Delta E/E$ of 20\% and an angular
resolution of around 15$^\circ$ for such events.

\section*{PINGU Science}

The primary scientific goal of PINGU is the observation of neutrino
oscillations using the atmospheric neutrino flux.  Several key
parameters will be measured by PINGU, including the mixing angles and
mass-squared splittings associated with both muon neutrino
disappearance and tau neutrino appearance, the octant of the mixing
angle $\thTT$, and the ordering of the neutrino mass eigenstates.

 \begin{figure}
   \begin{center}
     \includegraphics[width=5in]{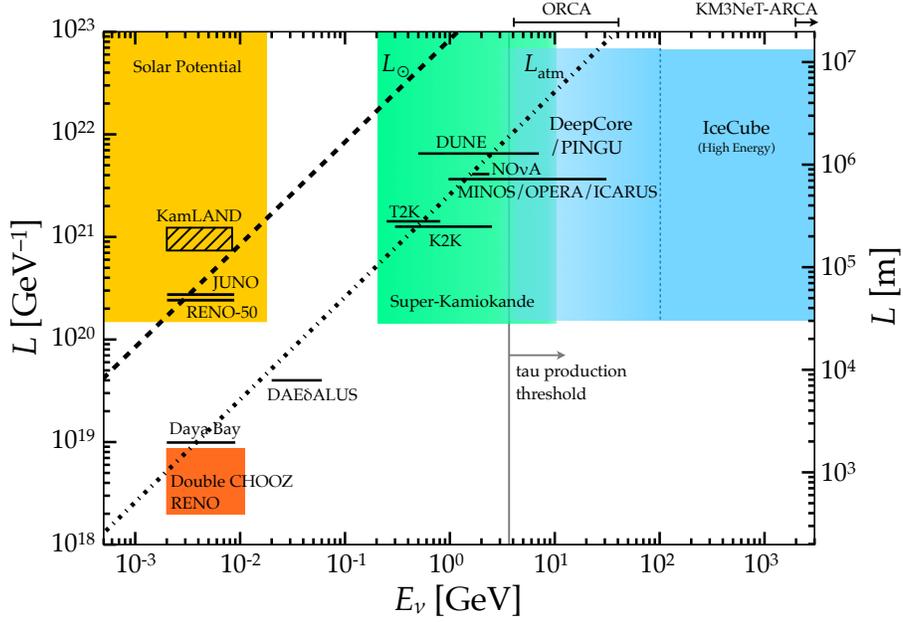}
   \end{center}
   \caption{
   Energy ranges and baselines of operational and planned
      neutrino oscillation experiments.  The diagonal  lines
      indicate the characteristic oscillation scales $L_{\astrosun}$ set by the solar mass-squared
     splitting $\Delta m^2_{21}$ (dashed) and $L_{\rm atm}$ set by
      the atmospheric mass-squared splitting  $\dmTT$ (dot-dashed).
      The 3.5~GeV threshold for $\tau$ lepton production in $\nutau$ CC events
      is shown by a vertical line.
      The energy ranges covered by the KM3NeT ORCA and ARCA detectors are indicated
      by bars above the plot for clarity.
      For Super-Kamiokande, ORCA, and PINGU, the upper end of the
      energy range is that at which the 
      $\numu$ energy resolution degrades because muons are no longer
      contained within the detector.  For IceCube and PINGU, this energy is
      marked by the vertical dashed line.
    }
   \label{fig:LvsE_ExecSumm}
 \end{figure}

 With neutrino path lengths through the Earth ranging up to 12,700~km,
 PINGU will observe the same oscillation phenomena at energies and
 baselines an order of magnitude larger than current and planned
 long-baseline neutrino beam experiments, as illustrated in
 Fig.~\ref{fig:LvsE_ExecSumm}.  PINGU thus complements accelerator and
 reactor neutrino experiments, as the different set of systematic
 uncertainties confronting PINGU and the weak impact of $\dcp$ on
 PINGU measurements will lend robustness to global determination of
 neutrino oscillation parameters.  Comparison of PINGU observations to those
 made by both currently running experiments such as T2K and NO$\nu$A
 and planned experiments such as DUNE, Hyper-Kamiokande, JUNO and
 KM3NeT / ORCA~\cite{ref:ARCAORCALoI} will also provide broad and
 model-independent potential for discovery of new physics. Finally, PINGU will
 have unprecedented sensitivity to tau neutrino
 appearance. Compared to the 180 charged current tau neutrino
 interactions observed in 2,806 days of Super-Kamiokande
 data~\cite{Abe:2012jj}, PINGU will be able to detect almost 
 3,000 such interactions every year.

The performance projections presented here are a summary of detailed
studies described in a more comprehensive document \cite{LOIv2inprep},
which will be available shortly.
They are based on full Monte Carlo simulations and detailed
reconstructions, including the full detector model developed over 10
years of experience operating the IceCube detector.  The full suite of
systematic uncertainties used for IceCube data analysis have been
taken into account in these studies.

\subsection*{Atmospheric Oscillation Measurements}

\begin{figure}
  \begin{center}
    \includegraphics[width=3.8in]{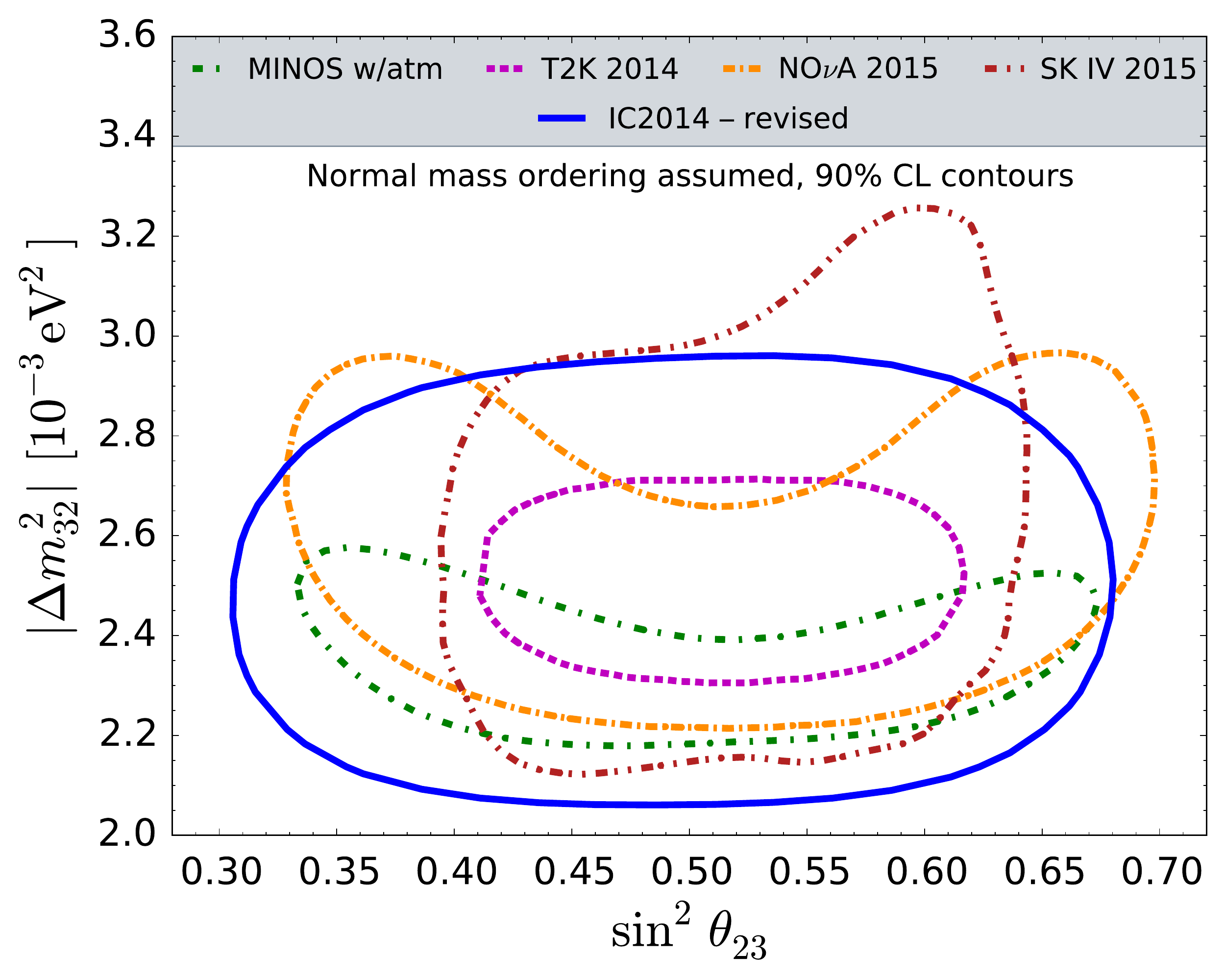}
  \end{center}
  \caption{Current measurements of the atmospheric mixing between the
    second and third mass eigenstates from atmospheric and
    long-baseline neutrino experiments.  Note that only the
    magnitude of the mass-squared splitting is known, not its sign.
   }
  \label{Fig:atm_mixing_current_ExecSumm}
\end{figure}

The ``atmospheric'' mixing between the second and third neutrino mass
eigenstates, which produced the first strong evidence that neutrinos
oscillate between flavors, is now the least well measured channel of
neutrino oscillation.  Current measurements of the atmospheric mixing
parameters $\sinsqTT$ and $\dmTT$ by IceCube~\cite{Aartsen:2014yll},
MINOS~\cite{Adamson:2013}, T2K~\cite{Abe:2014ugx},
NO$\nu$A~\cite{Adamson:2016xxw}, and
Super-Kamiokande~\cite{Nakahata:2015shm} are shown in
Fig.~\ref{Fig:atm_mixing_current_ExecSumm}.

\begin{figure}
\centering
\includegraphics[width=0.8\textwidth]{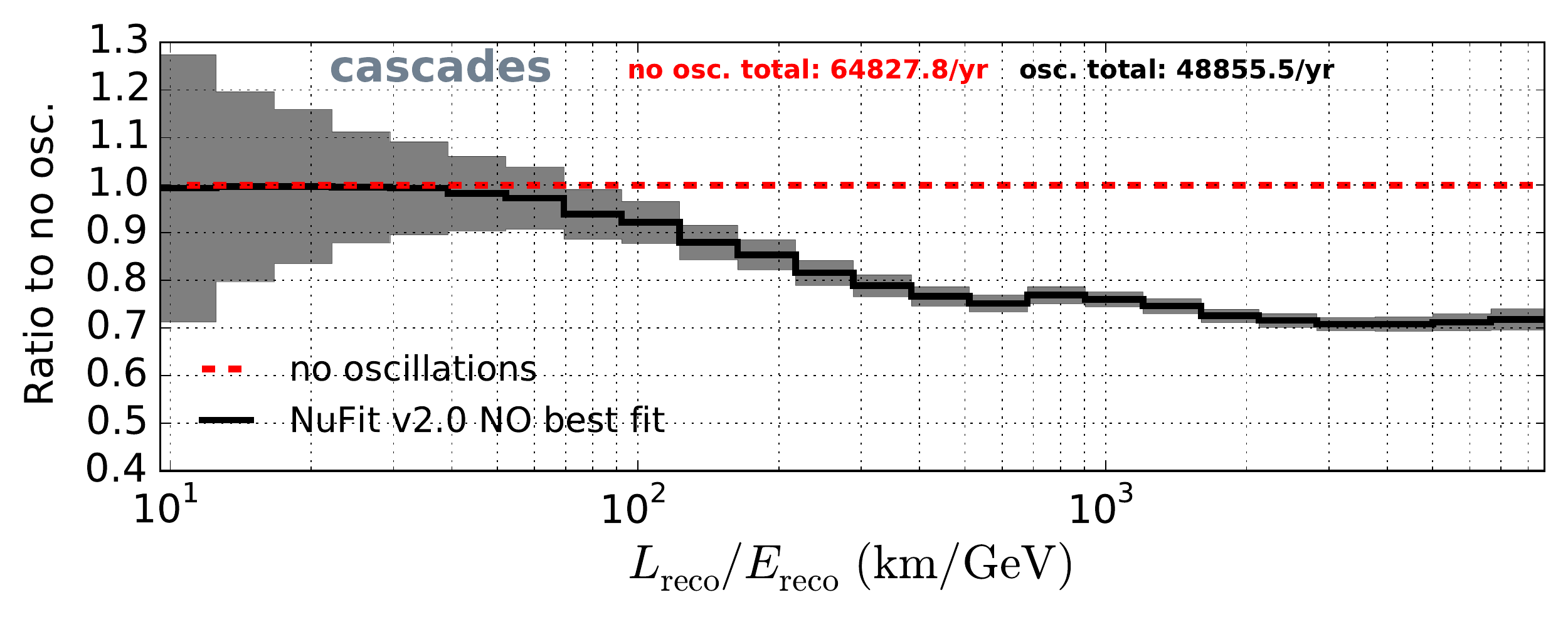}
\includegraphics[width=0.8\textwidth]{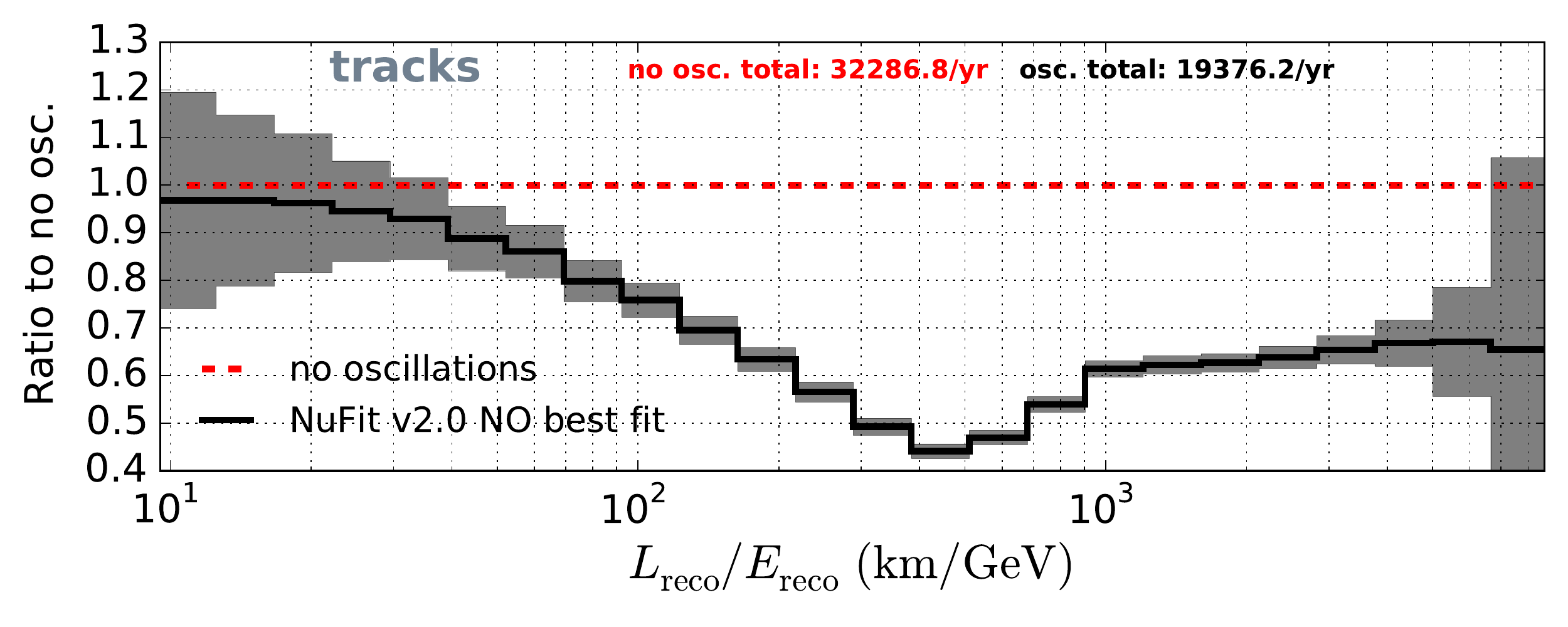}
\caption{The disappearance, caused by standard neutrino oscillations, that will be observed by PINGU in the cascade (top) and track (bottom) samples, as a function of the ratio of the reconstructed neutrino travel distance to its reconstructed energy. The gray bands show the sizes of the statistical uncertainties.}
\label{fig:LoverEplots}
\end{figure}

\begin{figure}
  \centering
  \subfigure[Normal neutrino mass ordering assumed.]{
        \includegraphics[scale=0.3]{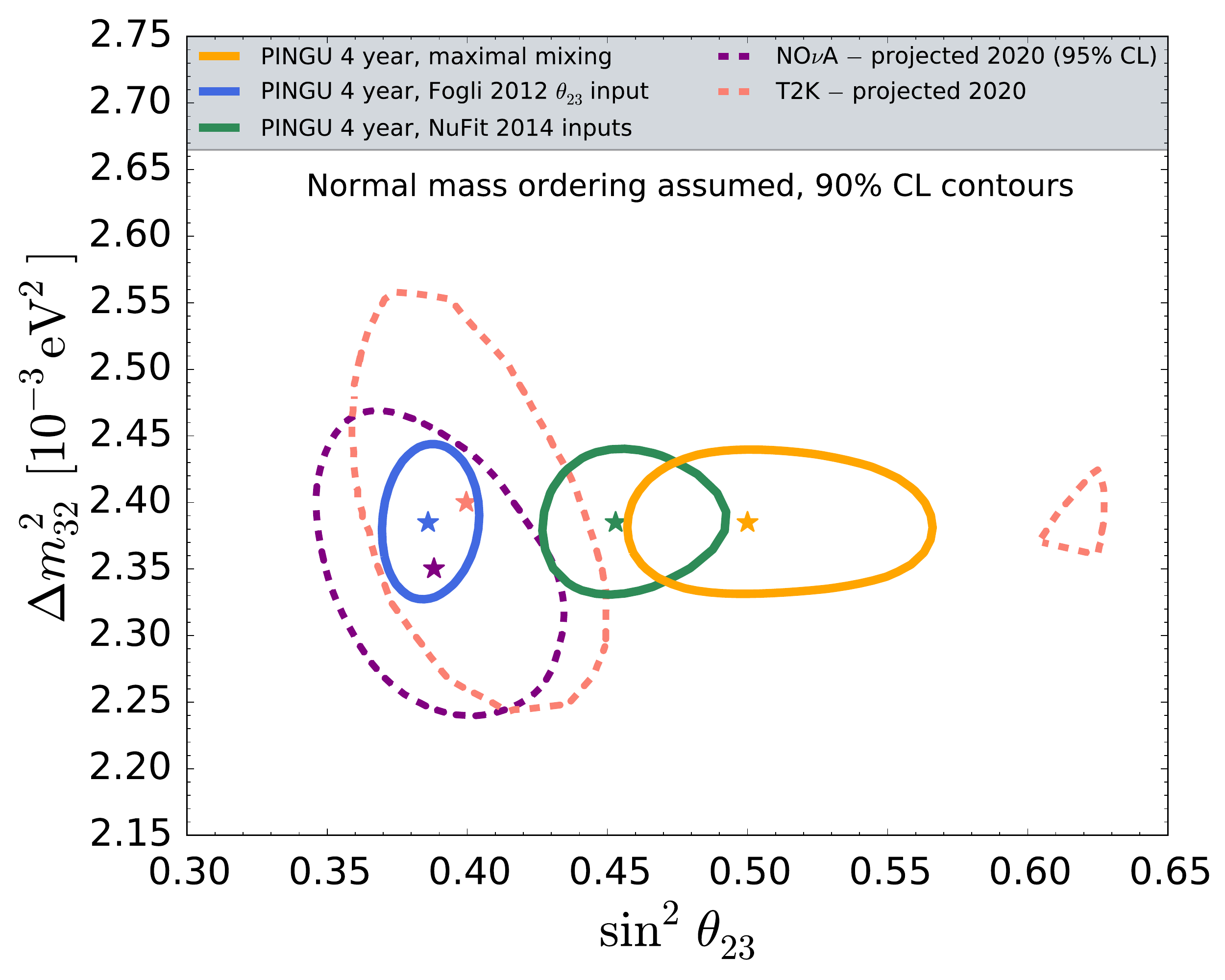}
        }
   \subfigure[Inverted neutrino mass ordering assumed.]{
        \includegraphics[scale=0.3]{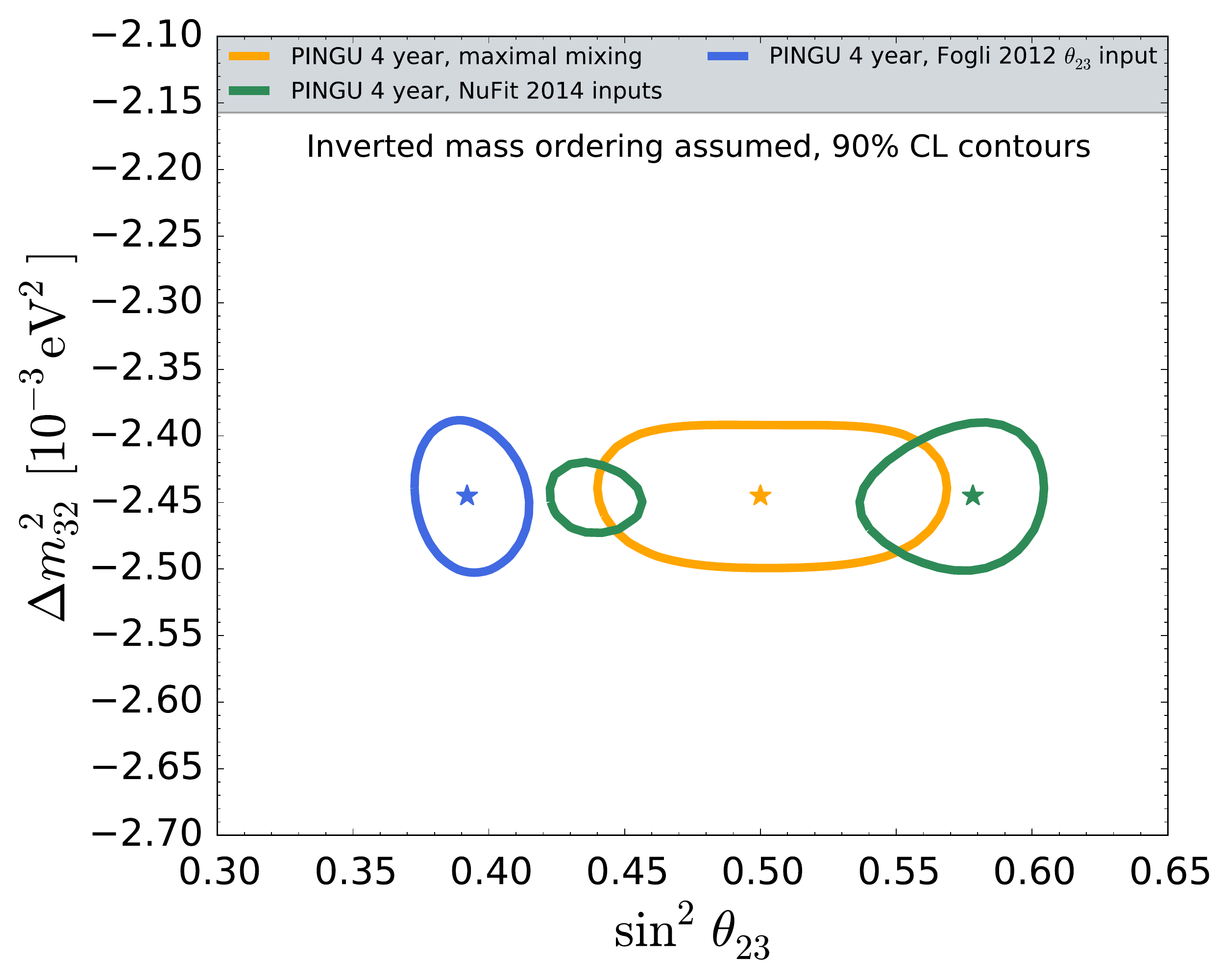}
        }
        \caption{The atmospheric neutrino
          oscillation contours are shown under assumptions of both the
          (a) normal and (b) inverted orderings.  Both orderings show
          the effect of different assumed true values: 
          the
          Fogli 2012~\cite{PhysRevD.86.013012} and NuFit
          2014~\cite{NuFIT20} global fits, and maximal mixing.  The normal ordering assumption
          includes projected sensitivities from NO$\nu$A (95\% CL, first
          octant only)~\cite{NOvA} and
          T2K~\cite{Abe:2014tzr} assuming $\dcp = 0$.  For NO$\nu$A,
          the second octant would be ruled out at 90\% CL under this assumption.}
  \label{fig:numu_disappearance_llhscan_ExecSumm}
\end{figure}

PINGU will measure the atmospheric parameters primarily through the
disappearance of $\numu$ from the atmospheric flux at energies above
5~GeV; Fig.~\ref{fig:LoverEplots} shows the disappearance that will be
observed by PINGU in the cascade and track samples as a function of
$L_{\mathrm{reco}}/E_{\mathrm{reco}}$, the reconstructed ratio of the
neutrino travel distance to its energy. With increased photocathode
density providing a lower energy threshold and significantly improved
event reconstruction compared to current IceCube
measurements~\cite{Aartsen:2014yll}, PINGU will determine these
parameters with precision comparable to or better than that expected
from current accelerator-based experiments
(Fig.~\ref{fig:numu_disappearance_llhscan_ExecSumm}), but at much
higher energies and over a range of very long baselines.  This will
provide world-class sensitivity to these parameters before the
next-generation long-baseline beam experiments such as
DUNE~\cite{Acciarri:2015uup} and Hyper-Kamiokande~\cite{Abe:2014oxa}
come online, as well as offering an important consistency check on the
standard oscillation paradigm and the potential for discovery of new
physics when higher precision measurements from next-generation
long-baseline instruments become available.

\subsection*{Maximal Mixing and the $\thTT$ Octant}

Current measurements of the mixing angle $\thTT$, which specifies the
relative amounts of the $\numu$ and $\nutau$ flavors in the third
neutrino mass eigenstate, suggest that the angle is close to
45$^\circ$ (corresponding to equal contributions from the two
flavors).  This possibility is known as ``maximal mixing'' and could
reflect a new fundamental symmetry.  If $\thTT$ is not exactly
45$^\circ$, determining its value and whether it is slightly more or
less than 45$^\circ$ (its ``octant'') is of great interest for
understanding the origin of neutrino masses and mixing
\cite{King:2015aea}.  In the simple two-flavor oscillation model,
values of $\thTT$ above and below 45$^\circ$ produce identical
transition probabilities.  However, this degeneracy is broken for
three-flavor oscillations in the presence of matter due to the large
value of $\theta_{13}$.

Neutrino beam experiments such as NO$\nu$A and T2K can probe the
$\thTT$ octant by comparison of $\nue$ appearance rates for neutrinos
and antineutrinos.  However, as the matter
effects at the energies and baselines of those experiments are
relatively weak, the sensitivity to the octant depends considerably on
the CP-violating parameter $\dcp$.  By contrast, PINGU will determine
the octant by comparison of $\numu \rightarrow \numu$ and
$\numu \rightarrow \nue$ transition probabilities for neutrinos and antineutrinos
passing through the Earth's core and
mantle~\cite{GonzalezGarcia:2004cu,Huber:2005ep,Barger:2012fx,Akhmedov:2012ah,Chatterjee:2013qus}.
The resonant matter effect on the conversion rates breaks the octant
degeneracy, and the value of $\dcp$ has little impact on PINGU
observations.

\begin{figure}[t]
  \centering
        \includegraphics[scale=.4]{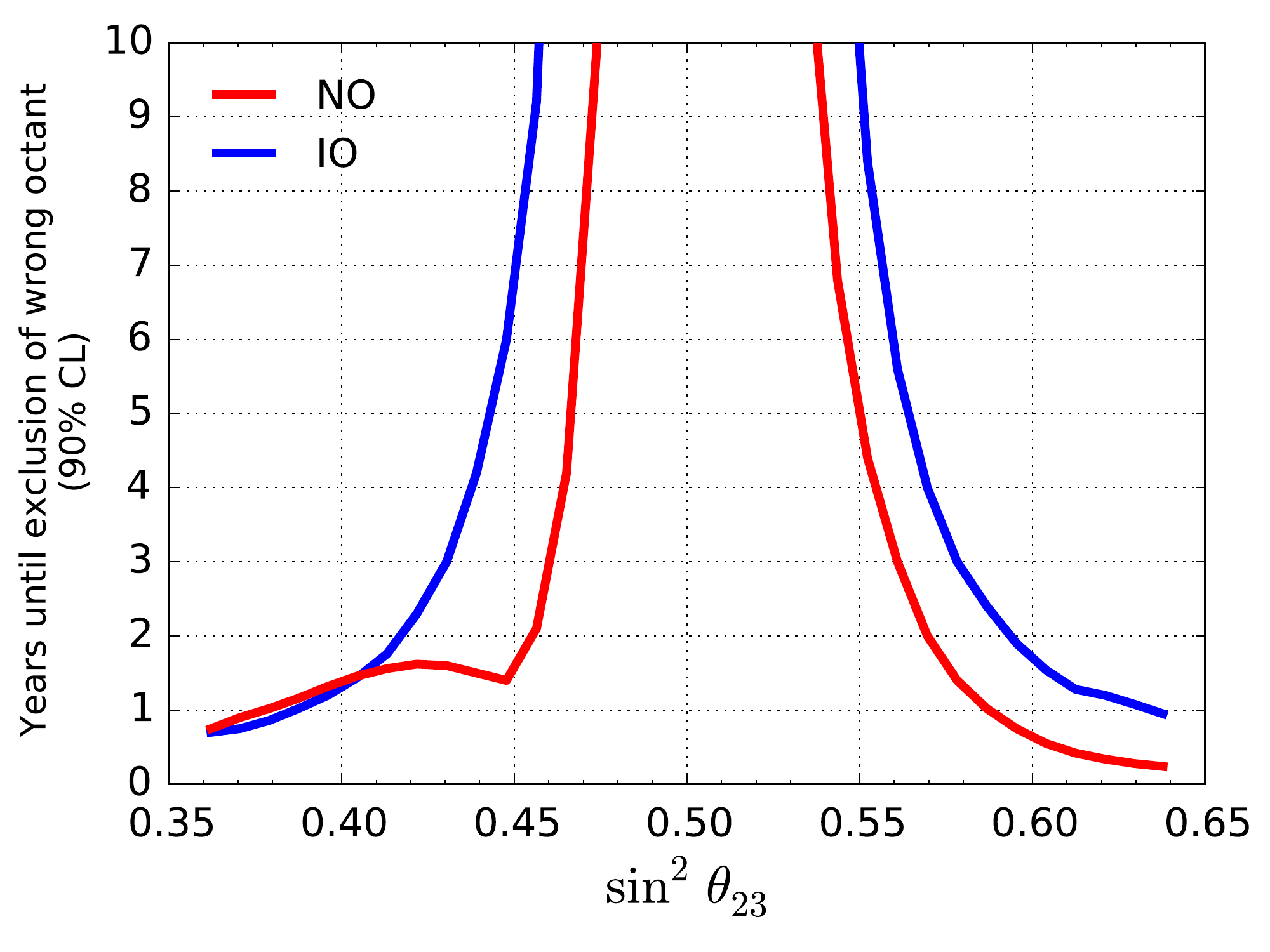}
        \caption{Amount of PINGU data required to determine the
          $\thTT$ octant (i.e., to exclude the wrong octant at 90\%
          C.L.), as a function of the true mass ordering and true
          value of $\sinsqTT$.  Sensitivity is lower if the ordering is inverted
         as the matter resonance affects antineutrinos rather than
         neutrinos.  The value of $\dcp$ has minimal impact and is
         assumed to be zero.}
        \label{fig:Octant_ExecSumm}
\end{figure}

The sensitivity of PINGU to the $\thTT$ octant is shown in
Fig.~\ref{fig:Octant_ExecSumm}.  If the neutrino mass ordering,
discussed in detail below, is
normal, PINGU's sensitivity is slightly better than expected for the combined T2K
and NO$\nu$A data sets~\cite{Agarwalla:2013ju}.  If the mass ordering
is inverted, PINGU is somewhat less sensitive than the long-baseline
experiments as the matter resonance affects antineutrinos.  In either
case, PINGU can determine the octant for a wide range of $\thTT$, and
for values close to maximal mixing PINGU data will be highly complementary to the long
baseline information due to the different sources of
degeneracy --- $\dcp$ for the beam experiments vs.~the mass ordering
for PINGU.



\subsection*{The Neutrino Mass Ordering}

The ordering of two of the three neutrino mass eigenstates,
$m(\nu_2) > m(\nu_1)$, is known from solar neutrino
measurements~\cite{ref:SNOPaper}, but we do not yet know whether
$\nu_3$ is heavier or lighter than the other two eigenstates.  This is
known as the {\it neutrino mass ordering} (NMO) question.  The case in
which $\nu_3$ is heavier is called the ``normal'' ordering (NO); if
$\nu_3$ is lighter, the ordering is ``inverted'' (IO).

In addition to its intrinsic interest, the ordering has deep
implications for the theoretical understanding of fundamental
interactions. Its measurement would assist in discriminating between
certain theoretical models at the GUT mass
scale~\cite{Mohapatra:2005wg}.  Experimentally, knowledge of the
ordering would positively impact ongoing and future research of other
crucial neutrino properties: the unknown NMO is a major ambiguity for
running or approved accelerator neutrino oscillation experiments with
sensitivity to leptonic CP
violation~\cite{Abe:2011ks,Messier:2013sfa,LBNE_Interim_Report,Hewett:2012ns}.
PINGU data are not highly sensitive to $\dcp$; if included as a
completely free nuisance parameter in the analysis, $\dcp$ reduces the
significance of the ordering determination by 10\%--20\% at most,
depending on the true values of $\dcp$ and $\thTT$.  In addition,
atmospheric neutrino data from PINGU or other proposed experiments
such as INO~\cite{Thakore:2013xqa} or
ORCA~\cite{Adrian-Martinez:2016fdl} in combination with existing neutrino beam
experiments and
proposed reactor experiments like JUNO~\cite{Djurcic:2015vqa} and RENO-50~\cite{Seo:2015yqp} 
provide synergistic inputs that can improve the
combined significance of the NMO determination beyond the purely
statistical addition of
results~\cite{Huber:2002rs,Winter:2013ema,Blennow:2013vta}.  PINGU's
determination of the NMO is thus highly complementary to other
experimental efforts, resolving possible degeneracies between the mass
ordering and CP violation and possibly increasing the precision with
which CP violation can be measured by long-baseline experiments.  In
addition, the determination of the NMO will influence the planning and 
interpretation of non-oscillation experiments (neutrinoless double
$\beta$ decay and $\beta$ decay) sensitive to the particle nature of the
neutrino (Dirac vs Majorana) and/or its absolute
mass~\cite{Feruglio:2002af}, and help to test popular
see-saw neutrino mass models and the related mechanism of leptogenesis
in the early universe~\cite{Fong:2013wr}.

\begin{figure}
  \centering
  \subfigure[Normal neutrino mass ordering assumed.]{
    \begin{overpic}[scale=0.3]{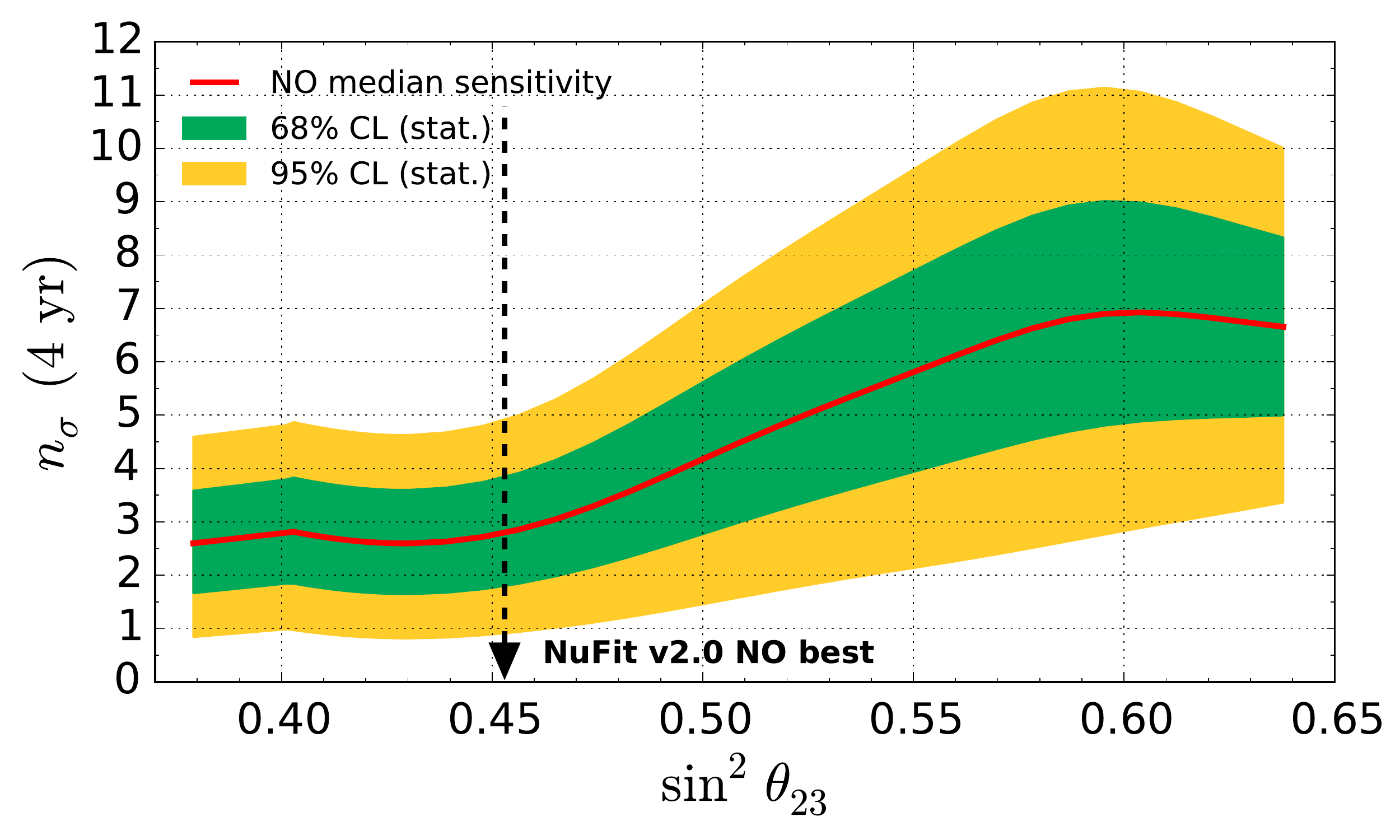}
    \end{overpic}
  }
  \subfigure[Inverted neutrino mass ordering assumed.]{
    \begin{overpic}[scale=0.3]{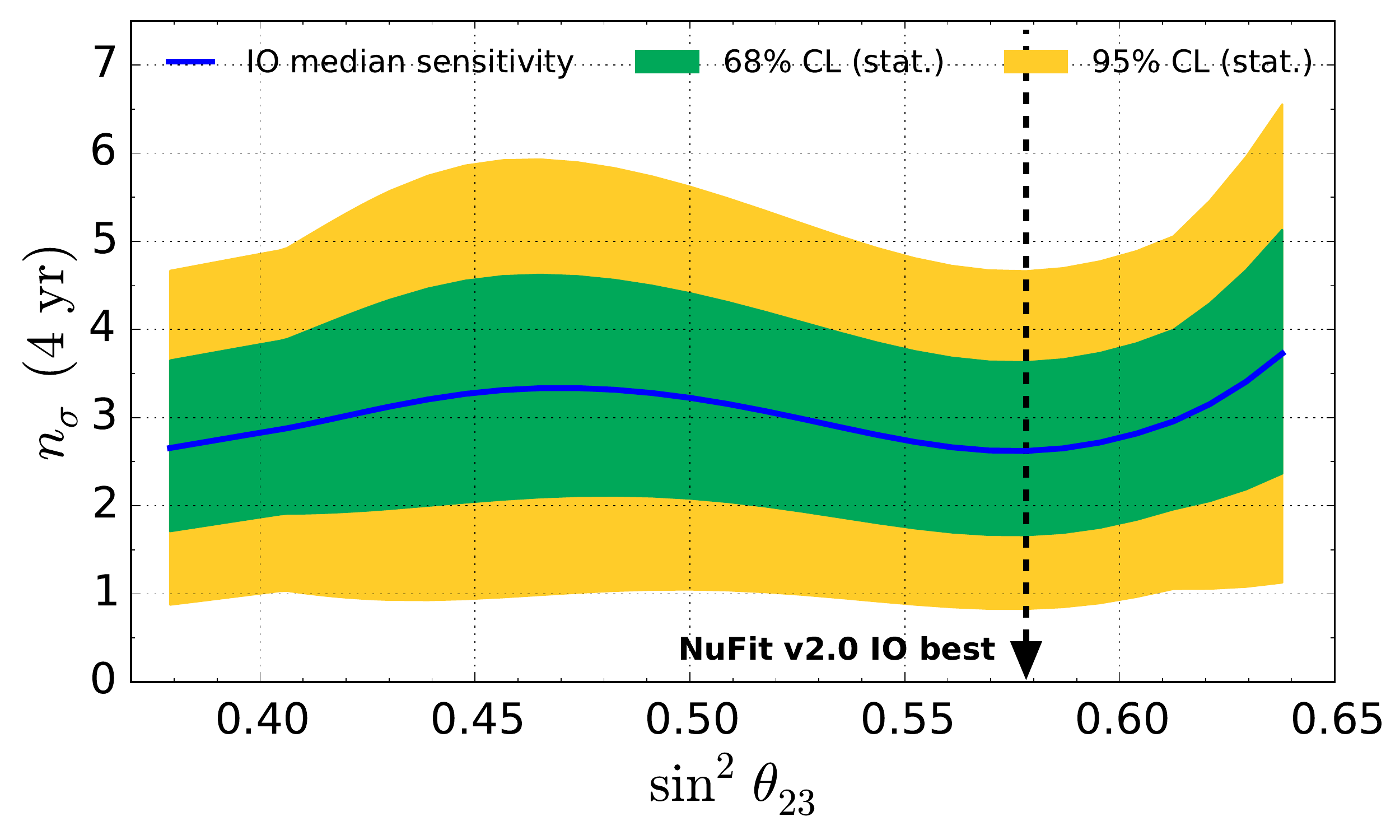}
    \end{overpic}
  }
  \caption{Expected significance with which the neutrino mass ordering
    will be determined using four years of data, as a function of the true
    value of $\sinsqTT$.  Solid red (NO) and blue (IO) lines show median
    significances, while the green and yellow bands indicate the range
    of significances obtained in 68\% and 95\% of hypothetical
    experiments.  The significance for determining the ordering when
    the true ordering is inverted is relatively insensitive to
    $\thTT$, while for the normal ordering large values of $\thTT$ are
    advantageous.  The range shown corresponds approximately to the current
    $3\sigma$ allowed region of $\thTT$; the global best-fit values from the
    NuFit group~\cite{NuFIT20} for both orderings are indicated by
    black arrows.}
  \label{fig:SigmaVsThetaTT_ExecSumm}
\end{figure}

With a neutrino energy threshold below \unit[5]{GeV}, PINGU will be
able to determine the NMO using the altered flavor composition of
atmospheric neutrinos that undergo Mikheyev-Smirnov-Wolfenstein
(MSW)~\cite{MSW-W,MSW-MS} and parametric~\cite{Akhmedov} oscillations
as they pass through the Earth.  At energies of approximately
\unit[5--20]{GeV}, the alteration of the oscillation patterns of both
$\numu$ and $\nue$ events is strong enough to enable PINGU to
determine whether the neutrino mass ordering is normal or inverted.
Given the current global best fit values of the oscillation
parameters, PINGU will determine the ordering with a median significance of
3$\sigma$ in approximately~\YrsToThreeSigmaSystLimited~years.  The
significance derived from any actual measurement is subject to large
statistical fluctuations, illustrated for PINGU in
Fig.~\ref{fig:SigmaVsThetaTT_ExecSumm}, so that multiple experimental
efforts to measure the ordering are required to guarantee it is
determined quickly.  For PINGU, the expected significance also 
depends strongly on the actual value of $\thTT$,
which is not well known.  The expectation
of~\YrsToThreeSigmaSystLimited~years to reach 3$\sigma$ significance
is conservative in the sense that PINGU's sensitivity to the NMO would
be greater in almost any region of the allowed parameter space of
$\thTT$ other than the current global best fit, as shown in
Fig.~\ref{fig:SigmaVsThetaTT_ExecSumm}.

\subsection*{Unitarity of the Neutrino Mixing Matrix}

\begin{figure}
  \begin{center}
    \begin{overpic}[scale=0.5]{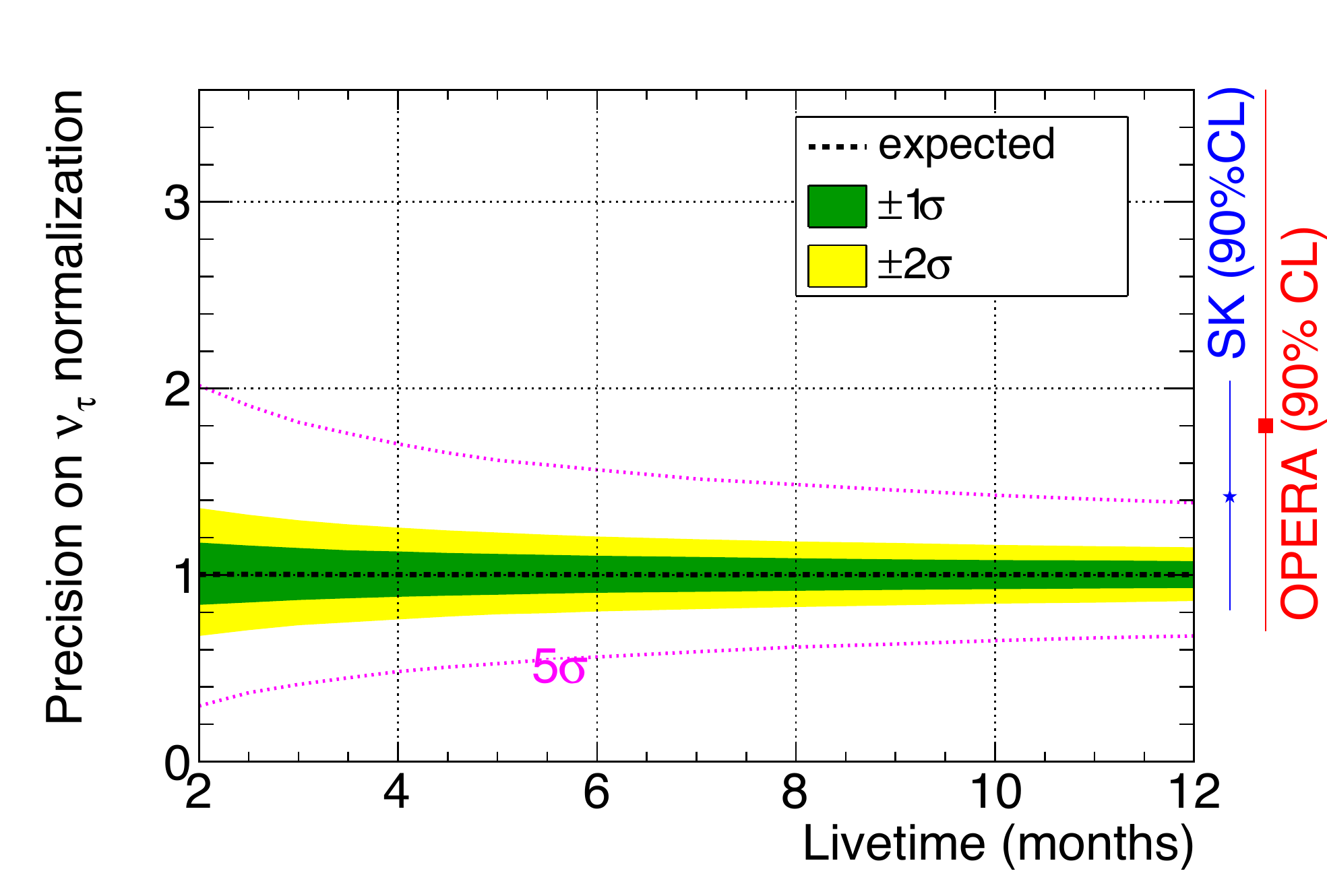}
    \end{overpic}
  \end{center}
  \caption{Precision with which the rate of $\nutau$ appearance can be
    measured, in terms of the PMNS expected rate, as a function of
    exposure (in months).  The true value is assumed to be 1.0 (the
    standard expectation) for illustration.  The expected
    $\pm 1\sigma$ and $\pm 2\sigma$ regions and $\pm 5\sigma$ limits
    are shown, as well as current measurements by
    Super-K~\cite{Abe:2012jj} and OPERA~\cite{Agafonova:2015jxn}.  
}
  \label{Fig:TauNeutrinoAppearance:NormalizationAndSignif_ExecSumm}
\end{figure}

In the standard neutrino oscillation picture, atmospheric $\numu$
disappearance arises primarily from $\numu \rightarrow \nutau$
oscillations.  However, in contrast to the CKM matrix in the quark
sector, the unitarity of the mixing between the three known neutrino
flavors has not been experimentally verified.  Many theories of
physics beyond the Standard Model include massive fermionic singlets
which could mix with neutrinos, expanding the standard $3\times 3$
PMNS neutrino mixing matrix into an extended $(3+N) \times (3+N)$
matrix and implying that the $3\times 3$ PMNS submatrix is
non-unitary.  The unitarity of PMNS mixing has only been tested at the
20\%-40\% level, primarily due to the lack of direct measurements of
$\nutau$ oscillations~\cite{Parke:2015goa}.  An extended mixing matrix
could either decrease or moderately increase the rate of $\nutau$
appearance relative to the Standard Model expectation.  Notably, both
the current measurements of $\nutau$ appearance somewhat exceed the
expected appearance rate, as shown in
Fig.~\ref{Fig:TauNeutrinoAppearance:NormalizationAndSignif_ExecSumm}.

The relatively high mass of the $\tau$ lepton greatly reduces
the interaction rate of $\nutau$ at low energies: current measurements
of $\nutau$ appearance rates are based on data sets including 180 
 and 5 $\nutau$ events in
Super-K~\cite{Abe:2012jj} and OPERAs~\cite{Agafonova:2015jxn},
respectively.  Tau
neutrino appearance on baselines comparable to the Earth's diameter
gives rise to large numbers of $\nutau$ with energies around
20~GeV, well above PINGU's energy threshold.  PINGU is expected to detect nearly
3,000 $\nutau$ CC interactions per year.  These $\nutau$ events can be
distinguished from the background of $\nue$ and $\numu$ CC and NC
events by their characteristic angular distribution and energy
spectrum, arising from their appearance via flavor oscillation at
specific $L_\nu/E_\nu$ (the ratio of the neutrino's path length
through the Earth to its energy).  This allows PINGU to measure
the rate of $\nutau$ appearance with a  precision of
better than 10\% with one year of data, as shown in
Fig.~\ref{Fig:TauNeutrinoAppearance:NormalizationAndSignif_ExecSumm},
providing a significantly more precise probe of PMNS matrix elements in the
$\numu$ and $\nutau$ rows than previous experiments.  The measurement
could either strengthen the 3-flavor model and the underlying
unitarity of its corresponding mixing, or point us in the direction of
new physics due to sterile neutrinos, non-standard interactions, or
other effects.


\subsection*{Additional PINGU Science: Dark Matter, Tomography and Supernovae}

By virtue of its GeV-scale neutrino energy threshold, PINGU will have
sensitivity to annihilations of  dark matter accreted by the Sun with mass as
low as 5~GeV.
In this neutrino energy regime, PINGU will also establish a new
experimental technique for direct tomographic measurement of the
Earth's composition through the faint imprint of the core's
proton-neutron ratio on neutrino oscillations
\cite{Rott:2015kwa,Winter:2015zwx}. Neutrino oscillation tomography
relies on the MSW effect, which depends on the electron
density. Seismic measurements by contrast are sensitive to the mass
density and have resulted in a very precise determination of the Earth
matter density profile, so the composition can be extracted from
comparison of the two measurements.  Although this technique is
affected by unknown neutrino physics, especially the octant of
$\thTT$, information regarding the Earth's composition can be
extracted with uncertainties in the oscillation physics and density
profile treated as nuisance parameters.  As global understanding of
the neutrino physics improves, more precise composition measurements
will be possible.

The increased density of instrumentation in PINGU compared to IceCube
and DeepCore will also enhance the observatory's sensitivity to bursts
of low energy ($\sim$ \unit[15]{MeV}) supernova neutrinos.  These
neutrinos are not detected individually, but rather observed as a
detector-wide increase in count rates due to the collective effect of
light deposited in the detector as the neutrino burst arrives
\cite{Abbasi:2011ss}.  Some information about the neutrino energy
spectrum can be obtained by comparing the rate at which immediately
neighboring DOMs detect light in close temporal coincidence,
indicative of a brighter neutrino event, to the overall count rate
\cite{Demiroers:2011am,Bruijn:2013ibl}.  The PINGU instrumentation
will provide an improvement in the sensitivity for detecting
supernovae of a factor of two and, due to the closer DOM spacing, a
factor of five in the precision of the measured average neutrino
energy \cite{LOIv2inprep}.

\section*{Cost, Schedule, and Logistics}


The 26 string configuration of PINGU substantially reduces costs in
several areas compared to the original 40 string configuration.
First, personnel costs associated with deployment are reduced significantly
by the elimination of the third drilling season.  Second, although the
number of optical modules increases slightly, other costs (cables,
fuel for the hot water drill, and logistical support) scale with the
number of holes and are cut by almost half.  Finally, the reduced
scope will allow us to refurbish the existing IceCube hot
water drill for reuse, rather than building a full replacement.

\begin{table}
\begin{center}
   \begin{tabular}{c|c|c} 
        & Cost (20 Strings) & Cost (26 Strings) \\ \hline \hline
      Drill refurbishment & \$5M  & \$5M  \\ 
      Deployment (labor) & \$5M  & \$5M  \\ 
      Instrumentation      & \$25M & \$33M \\
      Management \& other costs  & \$5M  & \$5M \\ \hline
     Total                        & \$39M & \$47M \\
      Fuel     & 146,000 gal & 190,000 gal\\ 

   \end{tabular}
   \caption{Summary costs in USD, excluding fuel and contingency, for
     construction of PINGU.  It is expected that non-US partners will
     provide the bulk of the instrumentation whose total cost is shown
     in the table. 
     Drill refurbishment and deployment  include the labor of the scientists and engineers associated
     with the hot water drill and string installation effort.
     Instrumentation costs include labor for module assembly, which
     contributes slightly over \$1M to the total.   Fuel 
     requirements for the hot water drill are provided as volumes due to
     uncertainties in the price of oil and the impact of the overland
     traverse on transport costs; recent costs are approximately
     \$20/gal.  
     \label{Tab:Costs_ExecSumm}}
\end{center}
\end{table}

Many components of the hot water drill used to install IceCube remain
available at the South Pole Station or in McMurdo Station, and reusing
them will greatly reduce the total project cost.  The formation of
bubbles in the re-frozen ice surrounding the optical modules is a
leading source of systematic uncertainty in IceCube data analyses.
The drill will be refurbished and a modified drill melting profile
will be used that will significantly reduce the quantity of
dissolved gases introduced into the detector region.  A water
filtration and degassing stage will be added to the drill to assist in
removing any remaining residual gases, thus limiting  bubble
formation.  The total cost of drill refurbishment and deployment
operations is approximately US\$10M.  The instrumentation for each
string costs approximately US\$1.2M; it is anticipated that the bulk
of the instrumentation would be provided by non-US participants.
Project management and other associated costs are expected to come to
an additional US\$5M.  A summary of costs is shown in
Table~\ref{Tab:Costs_ExecSumm}.

\begin{figure}
  \begin{center}
    \includegraphics[width=4.in]{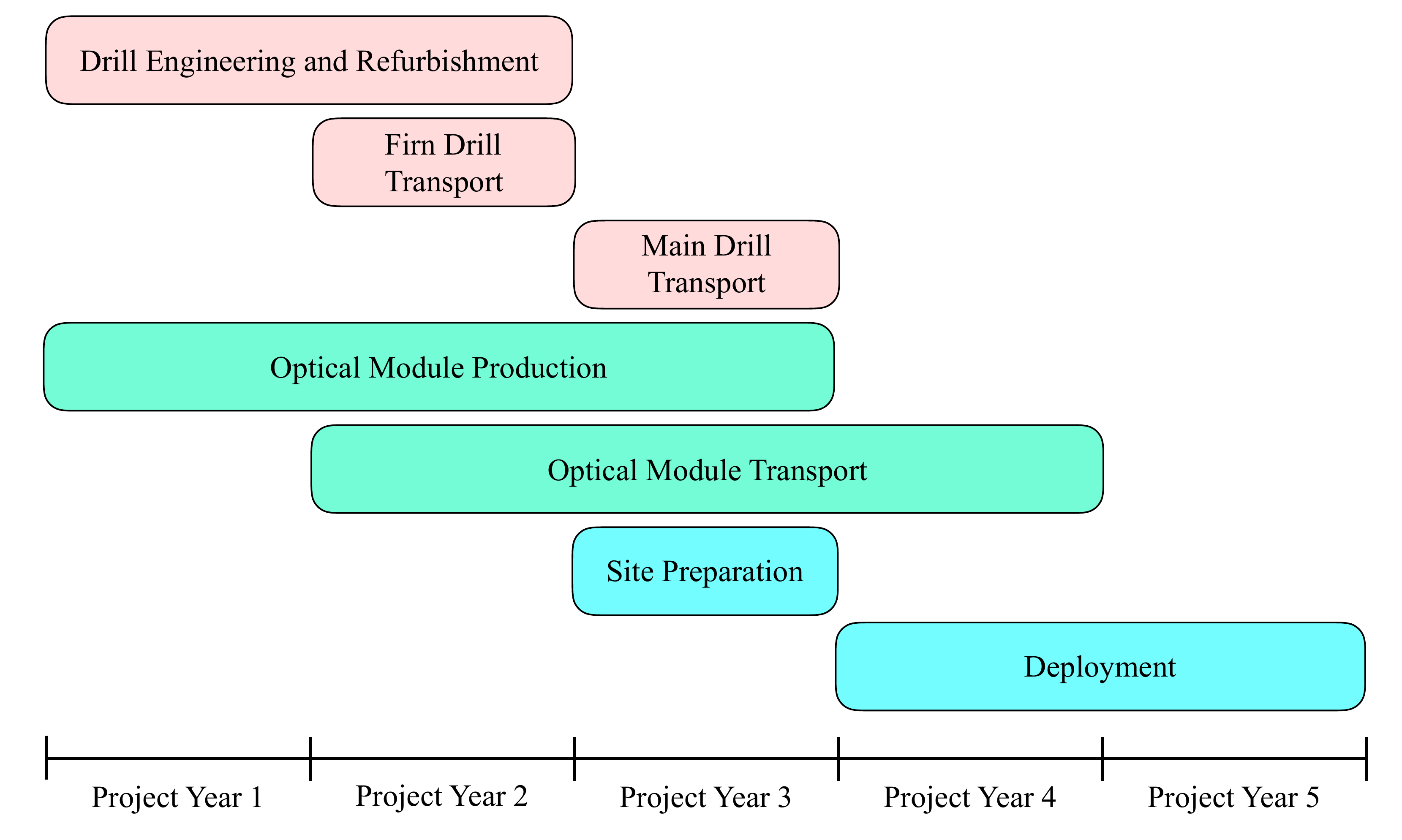}
  \end{center}
  \caption{Summary schedule for construction of PINGU.}
  \label{Fig:Schedule_ExecSumm}
\end{figure}

We anticipate that two years will be required for refurbishment and
improvement of the hot water drill.  Optical module assembly and
transportation to the South Pole would occur in parallel.  Once the
drill and optical modules are available at the South Pole, the full
PINGU array can be deployed in two seasons of activity.  Some
preparatory activity (snow compacting, firn drilling) would be
required in the preceding South Pole season to enable a prompt start to
deployment once the drill arrives.  A summary of the schedule is shown
in Figure~\ref{Fig:Schedule_ExecSumm}.

In contrast to the construction of the IceCube Observatory, for which
all cargo and fuel had to be airlifted to the South Pole Station,
nearly all materials required for PINGU construction would be
transported to the Pole via overland traverse.  In addition,
improvements in electronics design permit a substantial reduction in
power consumption by PINGU optical modules
 compared to IceCube DOMs.  Both of these advances will greatly
reduce the impact on Antarctic Program logistics, as well as reducing
costs.

\section*{Conclusion}


PINGU will be a world-class instrument for neutrino oscillation
physics exploring an energy and baseline range that cannot be probed
by long-baseline neutrino beam experiments.  PINGU will make a leading
measurement of the atmospheric neutrino oscillation parameters, test
the maximal mixing hypothesis, provide significantly improved
constraints on the unitarity of the Standard Model neutrino mixing matrix,
and determine the mass ordering with an expected significance of
3$\sigma$ within~\YrsToThreeSigmaSystLimited~years.
PINGU observations of high energy atmospheric neutrinos will be highly
complementary to existing and planned long-baseline and reactor
neutrino experiments, providing a robust validation with very
different systematic uncertainties as well as sensitivity to potential
new physics.  PINGU will also extend IceCube's reach in
searches for dark matter annihilation to low mass particles,
increase our sensitivity to neutrino bursts from
supernovae, and provide a first-ever tomographic probe of the Earth's
core.

Building on prior experience with IceCube and DeepCore, the risks
associated with instrumentation design, drilling, and deployment are 
well understood and proven to be manageable.  Likewise, the estimated
cost is well grounded in knowledge gained in the design
and construction of IceCube.  The performance projections shown here
are based on full detector simulation and reconstruction algorithms
informed by a decade of experience operating IceCube.  Moreover, there
is potential for further improvements in the future using a detector based on multi-PMT DOMs.

\resetlinenumber
\section*{Acknowledgements}

We acknowledge the support from the following agencies: U.S. National
Science Foundation-Office of Polar Programs, U.S. National Science
Foundation-Physics Division, University of Wisconsin Alumni Research
Foundation, the Grid Laboratory Of Wisconsin (GLOW) grid
infrastructure at the University of Wisconsin - Madison, the Open
Science Grid (OSG) grid infrastructure; U.S. Department of Energy, and
National Energy Research Scientific Computing Center, the Louisiana
Optical Network Initiative (LONI) grid computing resources; Natural
Sciences and Engineering Research Council of Canada, WestGrid and
Compute/Calcul Canada; Swedish Research Council, Swedish Polar
Research Secretariat, Swedish National Infrastructure for Computing
(SNIC), and Knut and Alice Wallenberg Foundation, Sweden; German
Ministry for Education and Research (BMBF), Deutsche
Forschungsgemeinschaft (DFG), Helmholtz Alliance for Astroparticle
Physics (HAP), Research Department of Plasmas with Complex
Interactions (Bochum), Germany; Fund for Scientific Research
(FNRS-FWO), FWO Odysseus programme, Flanders Institute to encourage
scientific and technological research in industry (IWT), Belgian
Federal Science Policy Office (Belspo); Science and Technology
Facilities Council (STFC) and University of Oxford, United
Kingdom; Marsden Fund, New Zealand; Australian Research Council; Japan
Society for Promotion of Science (JSPS); the Swiss National Science
Foundation (SNSF), Switzerland; National Research Foundation of Korea
(NRF); Villum Fonden, Danish National Research Foundation (DNRF),
Denmark.

\newpage
\resetlinenumber
\bibliography{biblio}  

\end{document}

%% file: NewCommands.tex
%
%
%

%
%
%

%
%
\providecommand{\YrsToThreeSigmaSystLimited}{5}

%
%
%

%
%

\providecommand{\dmTT}{\Delta m^2_{\rm 32}}

\providecommand{\thTT}{\theta_{\rm 23}}

\providecommand{\sinsqTT}{\sin^{2}(\theta_{\rm 23})}

\providecommand{\dcp}{\delta_{\rm CP}}

\providecommand{\nue}{\nu_{\rm e}}
\providecommand{\numu}{\nu_\mu}
\providecommand{\nutau}{\nu_\tau}

%% file: AuthorList.tex
\author[Adelaide]{M.~G.~Aartsen} 
\author[Munich]{K.~Abraham} 
\author[Zeuthen]{M.~Ackermann} 
\author[Christchurch]{J.~Adams} 
\author[BrusselsLibre]{J.~A.~Aguilar} 
\author[MadisonPAC]{M.~Ahlers} 
\author[StockholmOKC]{M.~Ahrens} 
\author[Erlangen]{D.~Altmann} 
\author[Marquette]{K.~Andeen} 
\author[PennPhys]{T.~Anderson} 
% ...
\address[Adelaide]{Department of Physics, University of Adelaide, Adelaide, 5005, Australia}
% ...
\address[BrusselsLibre]{Universit\'e Libre de Bruxelles, Science Faculty CP230, B-1050 Brussels, Belgium}
% ...
\address[Christchurch]{Dept.~of Physics and Astronomy, University of Canterbury, Private Bag 4800, Christchurch, New Zealand}
% ...
\address[Erlangen]{Erlangen Centre for Astroparticle Physics, Friedrich-Alexander-Universit\"at Erlangen-N\"urnberg, D-91058 Erlangen, Germany}
% ...
\address[MadisonPAC]{Dept.~of Physics and Wisconsin IceCube Particle Astrophysics Center, University of Wisconsin, Madison, WI 53706, USA}
% ...
\address[Marquette]{Department of Physics, Marquette University, Milwaukee, WI, 53201, USA}
% ...
\address[Munich]{Physik-Department, Technische Universit\"at M\"unchen, D-85748 Garching, Germany}
% ...
\address[StockholmOKC]{Oskar Klein Centre and Dept.~of Physics, Stockholm University, SE-10691 Stockholm, Sweden}
% ...
\address[PennPhys]{Dept.~of Physics, Pennsylvania State University, University Park, PA 16802, USA}
% ...
\address[Zeuthen]{DESY, D-15735 Zeuthen, Germany}

%% file: ShortSummary.bbl
\providecommand{\href}[2]{#2}\begingroup\raggedright\begin{thebibliography}{10}

\bibitem{Fukuda:1998mi}
{\bf Super-Kamiokande} Collaboration, Y.~Fukuda {\em et~al.}, ``{Evidence for
  oscillation of atmospheric neutrinos},'' {\em Phys. Rev. Lett.} {\bf 81}
  (1998) 1562--1567, \href{http://xxx.lanl.gov/abs/arXiv:hep-ex/9807003}{{\tt
  arXiv:hep-ex/9807003}}.

\bibitem{Ahmad:2001an}
{\bf SNO} Collaboration, Q.~R. Ahmad {\em et~al.}, ``Measurement of the rate of
  $\nu_e + d \to p + p + e^-$ interactions produced by $^8${B} solar neutrinos
  at the {Sudbury Neutrino Observatory},'' {\em Phys. Rev. Lett.} {\bf 87}
  (2001) 071301, \href{http://xxx.lanl.gov/abs/arXiv:nucl-ex/0106015}{{\tt
  arXiv:nucl-ex/0106015}}.

\bibitem{Aartsen:2013pza}
{\bf IceCube} Collaboration, M.~G. Aartsen {\em et~al.}, ``{Evidence for
  high-energy extraterrestrial neutrinos at the IceCube detector},'' {\em
  Science} {\bf 342} (2013) 1242856,
  \href{http://xxx.lanl.gov/abs/arXiv:1311.5238}{{\tt arXiv:1311.5238}}.

\bibitem{Aartsen:2013jza}
{\bf IceCube} Collaboration, M.~Aartsen {\em et~al.}, ``{Measurement of
  atmospheric neutrino oscillations with IceCube},'' {\em Phys.~Rev.~Lett.}
  {\bf 111} (2013) 081801, \href{http://xxx.lanl.gov/abs/arXiv:1305.3909}{{\tt
  arXiv:1305.3909}}.

\bibitem{Aartsen:2014yll}
{\bf IceCube} Collaboration, M.~Aartsen {\em et~al.}, ``{Determining neutrino
  oscillation parameters from atmospheric muon neutrino disappearance with
  three years of IceCube DeepCore data},'' {\em Phys. Rev.} {\bf D91} (2015)
  072004, \href{http://xxx.lanl.gov/abs/arXiv:1410.7227}{{\tt
  arXiv:1410.7227}}.

\bibitem{Aartsen:2012kia}
{\bf IceCube} Collaboration, M.~Aartsen {\em et~al.}, ``{Search for dark matter
  annihilations in the Sun with the 79-string IceCube detector},'' {\em
  Phys.~Rev.~Lett.} {\bf 110} (2013) 131302,
  \href{http://xxx.lanl.gov/abs/arXiv:1212.4097}{{\tt arXiv:1212.4097}}.

\bibitem{Aartsen:2013rt}
{\bf IceCube} Collaboration, M.~Aartsen {\em et~al.}, ``{Measurement of South
  Pole ice transparency with the IceCube LED calibration system},'' {\em
  Nucl.~Instrum.~Meth.} {\bf A711} (2013) 73--89,
  \href{http://xxx.lanl.gov/abs/arXiv:1301.5361}{{\tt arXiv:1301.5361}}.

\bibitem{Aartsen:2014oha}
{\bf IceCube PINGU} Collaboration, M.~G. Aartsen {\em et~al.}, ``{Letter of
  Intent: the Precision IceCube Next Generation Upgrade (PINGU)},''
  \href{http://xxx.lanl.gov/abs/arXiv:1401.2046}{{\tt arXiv:1401.2046}}.

\bibitem{Abbasi:2008aa}
{\bf IceCube} Collaboration, R.~Abbasi {\em et~al.}, ``{The IceCube data
  acquisition system: signal capture, digitization, and timestamping},'' {\em
  Nucl. Instrum. Meth.} {\bf A601} (2009) 294--316,
  \href{http://xxx.lanl.gov/abs/arXiv:0810.4930}{{\tt arXiv:0810.4930}}.

\bibitem{Abbasi:2010vc}
{\bf IceCube} Collaboration, R.~Abbasi {\em et~al.}, ``{Calibration and
  characterization of the IceCube photomultiplier tube},'' {\em Nucl. Instrum.
  Meth.} {\bf A618} (2010) 139--152,
  \href{http://xxx.lanl.gov/abs/arXiv:1002.2442}{{\tt arXiv:1002.2442}}.

\bibitem{Classen:2013xma}
{\bf KM3NeT} Collaboration, L.~Classen and O.~Kalekin, ``{Status of the PMT
  development for KM3NeT},'' {\em Nucl.~Instrum.~Meth.} {\bf A725} (2013)
  155--157.

\bibitem{Adrian-Martinez:2014vja}
{\bf KM3NeT} Collaboration, S.~Adrian-Martinez {\em et~al.}, ``{Deep sea tests
  of a prototype of the KM3NeT digital optical module},'' {\em Eur.~Phys.~J.}
  {\bf C74} (2014) 3056.

\bibitem{ref:ARCAORCALoI}
{\bf KM3NeT} Collaboration, S.~Adri{\'a}n-Mart{\'i}nez {\em et~al.}, ``{Letter
  of intent for KM3NeT 2.0},'' {\em J. Phys. G} {\bf 43} (2016) 084001,
  \href{http://xxx.lanl.gov/abs/arXiv:1601.07459}{{\tt arXiv:1601.07459}}.

\bibitem{Abe:2012jj}
{\bf Super-Kamiokande} Collaboration, K.~Abe {\em et~al.}, ``{Evidence for the
  appearance of atmospheric tau neutrinos in Super-Kamiokande},'' {\em Phys.
  Rev. Lett.} {\bf 110} (2013) 181802,
  \href{http://xxx.lanl.gov/abs/arXiv:1206.0328}{{\tt arXiv:1206.0328}}.

\bibitem{LOIv2inprep}
{\bf IceCube-Gen2} Collaboration. In preparation.

\bibitem{Adamson:2013}
{\bf MINOS} Collaboration, P.~Adamson {\em et~al.}, ``{Measurement of neutrino
  and antineutrino oscillations using beam and atmospheric data in MINOS},''
  {\em Phys.~Rev.~Lett.} {\bf 110} (2013) 251801,
  \href{http://xxx.lanl.gov/abs/arXiv:1304.6335}{{\tt arXiv:1304.6335}}.

\bibitem{Abe:2014ugx}
{\bf T2K} Collaboration, K.~Abe {\em et~al.}, ``{Precise measurement of the
  neutrino mixing parameter $\theta_{23}$ from muon neutrino disappearance in
  an off-axis beam},'' {\em Phys.Rev.Lett.} {\bf 112} (2014) 181801,
  \href{http://xxx.lanl.gov/abs/arXiv:1403.1532}{{\tt arXiv:1403.1532}}.

\bibitem{Adamson:2016xxw}
{\bf NOvA} Collaboration, P.~Adamson {\em et~al.}, ``{First measurement of
  muon-neutrino disappearance in NOvA},'' {\em Phys. Rev.} {\bf D93} (2016)
  051104, \href{http://xxx.lanl.gov/abs/arXiv:1601.05037}{{\tt
  arXiv:1601.05037}}.

\bibitem{Nakahata:2015shm}
{\bf Super-Kamiokande} Collaboration, M.~Nakahata, ``{Recent results from
  Super-Kamiokande},'' {\em PoS} {\bf NEUTEL2015} (2015) 009.

\bibitem{Acciarri:2015uup}
{\bf DUNE} Collaboration, R.~Acciarri {\em et~al.}, ``{Long-Baseline Neutrino
  Facility (LBNF) and Deep Underground Neutrino Experiment (DUNE)},''
  \href{http://xxx.lanl.gov/abs/arXiv:1512.06148}{{\tt arXiv:1512.06148}}.

\bibitem{Abe:2014oxa}
{\bf Hyper-Kamiokande Working Group} Collaboration, K.~Abe {\em et~al.}, ``{A
  long baseline neutrino oscillation experiment using J-PARC neutrino beam and
  Hyper-Kamiokande},'' \href{http://xxx.lanl.gov/abs/arXiv:1412.4673}{{\tt
  arXiv:1412.4673}}.

\bibitem{PhysRevD.86.013012}
G.~L. Fogli, E.~Lisi, A.~Marrone, D.~Montanino, A.~Palazzo, and A.~M. Rotunno,
  ``Global analysis of neutrino masses, mixings, and phases: Entering the era
  of leptonic {CP} violation searches,'' {\em Phys.~Rev.} {\bf D86} (2012)
  013012.

\bibitem{NuFIT20}
M.~Gonzalez-Garcia, M.~Maltoni, and T.~Schwetz, ``Updated fit to three neutrino
  mixing: status of leptonic {CP} violation,'' {\em Journal of High Energy
  Physics} {\bf 2014} (2014) 52.

\bibitem{NOvA}
``{NOvA} plots and figures.''
  http://www-nova.fnal.gov/plots\_and\_figures/plot\_and\_figures.html.
\newblock Accessed: 2015.

\bibitem{Abe:2014tzr}
{\bf T2K} Collaboration, K.~Abe {\em et~al.}, ``{Neutrino oscillation physics
  potential of the T2K experiment},'' {\em PTEP} {\bf 2015} (2015) 043C01,
  \href{http://xxx.lanl.gov/abs/arXiv:1409.7469}{{\tt arXiv:1409.7469}}.

\bibitem{King:2015aea}
S.~F. King, ``{Models of Neutrino Mass, Mixing and {CP} Violation},'' {\em J.
  Phys.} {\bf G42} (2015) 123001,
  \href{http://xxx.lanl.gov/abs/arXiv:1510.02091}{{\tt arXiv:1510.02091}}.

\bibitem{GonzalezGarcia:2004cu}
M.~C. Gonzalez-Garcia, M.~Maltoni, and A.~{\relax Yu}. Smirnov, ``{Measuring
  the deviation of the 2-3 lepton mixing from maximal with atmospheric
  neutrinos},'' {\em Phys. Rev.} {\bf D70} (2004) 093005,
  \href{http://xxx.lanl.gov/abs/arXiv:hep-ph/0408170}{{\tt
  arXiv:hep-ph/0408170}}.

\bibitem{Huber:2005ep}
P.~Huber, M.~Maltoni, and T.~Schwetz, ``{Resolving parameter degeneracies in
  long-baseline experiments by atmospheric neutrino data},'' {\em Phys. Rev.}
  {\bf D71} (2005) 053006,
  \href{http://xxx.lanl.gov/abs/arXiv:hep-ph/0501037}{{\tt
  arXiv:hep-ph/0501037}}.

\bibitem{Barger:2012fx}
V.~Barger, R.~Gandhi, P.~Ghoshal, S.~Goswami, D.~Marfatia, S.~Prakash, S.~K.
  Raut, and S.~U. Sankar, ``{Neutrino mass hierarchy and octant determination
  with atmospheric neutrinos},'' {\em Phys. Rev. Lett.} {\bf 109} (2012)
  091801, \href{http://xxx.lanl.gov/abs/arXiv:1203.6012}{{\tt
  arXiv:1203.6012}}.

\bibitem{Akhmedov:2012ah}
E.~K. Akhmedov, S.~Razzaque, and A.~{\relax Yu}. Smirnov, ``{Mass hierarchy,
  2-3 mixing and {CP}-phase with huge atmospheric neutrino detectors},'' {\em
  JHEP} {\bf 02} (2013) 082,
  \href{http://xxx.lanl.gov/abs/arXiv:1205.7071}{{\tt arXiv:1205.7071}}.
  [Erratum: JHEP07,026(2013)].

\bibitem{Chatterjee:2013qus}
A.~Chatterjee, P.~Ghoshal, S.~Goswami, and S.~K. Raut, ``{Octant sensitivity
  for large $\theta_{13}$ in atmospheric and long baseline neutrino
  experiments},'' {\em JHEP} {\bf 06} (2013) 010,
  \href{http://xxx.lanl.gov/abs/arXiv:1302.1370}{{\tt arXiv:1302.1370}}.

\bibitem{Agarwalla:2013ju}
S.~K. Agarwalla, S.~Prakash, and S.~U. Sankar, ``{Resolving the octant of
  $\theta_{23}$ with T2K and NOvA},'' {\em JHEP} {\bf 07} (2013) 131,
  \href{http://xxx.lanl.gov/abs/arXiv:1301.2574}{{\tt arXiv:1301.2574}}.

\bibitem{ref:SNOPaper}
{\bf SNO} Collaboration, B.~Aharmim {\em et~al.}, ``{Electron energy spectra,
  fluxes, and day-night asymmetries of $^8${B} solar neutrinos from
  measurements with NaCl dissolved in the heavy-water detector at the Sudbury
  Neutrino Observatory},'' {\em Phys. Rev.} {\bf C72} (2005) 055502,
  \href{http://xxx.lanl.gov/abs/arXiv:nucl-ex/0502021}{{\tt
  arXiv:nucl-ex/0502021}}.

\bibitem{Mohapatra:2005wg}
R.~Mohapatra, S.~Antusch, K.~Babu, G.~Barenboim, M.-C. Chen, {\em et~al.},
  ``{Theory of neutrinos: a white paper},'' {\em Rept.~Prog.~Phys.} {\bf 70}
  (2007) 1757--1867, \href{http://xxx.lanl.gov/abs/arXiv:hep-ph/0510213}{{\tt
  arXiv:hep-ph/0510213}}.

\bibitem{Abe:2011ks}
{\bf T2K} Collaboration, K.~Abe {\em et~al.}, ``{The T2K Experiment},'' {\em
  Nucl.~Instrum.~Meth.} {\bf A659} (2011) 106--135,
  \href{http://xxx.lanl.gov/abs/arXiv:1106.1238}{{\tt arXiv:1106.1238}}.

\bibitem{Messier:2013sfa}
{\bf NOvA} Collaboration, M.~D. Messier, ``{Extending the NOvA Physics
  Program},'' in {\em {Report of the Community Summer Study 2013: Snowmass on
  the Mississippi, Minneapolis, MN, USA}}, 2013.
\newblock \href{http://xxx.lanl.gov/abs/arXiv:1308.0106}{{\tt
  arXiv:1308.0106}}.

\bibitem{LBNE_Interim_Report}
{\bf LBNE} Collaboration, T.~Akiri {\em et~al.}, ``{The 2010 interim report of
  the Long-Baseline Neutrino Experiment Collaboration physics working
  groups},'' \href{http://xxx.lanl.gov/abs/arXiv:1110.6249}{{\tt
  arXiv:1110.6249}}.

\bibitem{Hewett:2012ns}
H.~Weerts {\em et~al.}, eds., {\em {Fundamental Physics at the Intensity
  Frontier}}, Report of the workshop held December 2011 in Rockville, MD, 2012.
\newblock \href{http://xxx.lanl.gov/abs/arXiv:1205.2671}{{\tt
  arXiv:1205.2671}}.

\bibitem{Thakore:2013xqa}
T.~Thakore, A.~Ghosh, S.~Choubey, and A.~Dighe, ``{The reach of INO for
  atmospheric neutrino oscillation parameters},'' {\em J.~High Energy Phys.}
  {\bf 1305} (2013) 058, \href{http://xxx.lanl.gov/abs/arXiv:1303.2534}{{\tt
  arXiv:1303.2534}}.

\bibitem{Adrian-Martinez:2016fdl}
{\bf KM3Net} Collaboration, S.~Adrian-Martinez {\em et~al.}, ``{Letter of
  Intent for KM3NeT2.0},'' \href{http://xxx.lanl.gov/abs/arXiv:1601.07459}{{\tt
  arXiv:1601.07459}}.

\bibitem{Djurcic:2015vqa}
{\bf JUNO} Collaboration, Z.~Djurcic {\em et~al.}, ``{JUNO Conceptual Design
  Report},'' \href{http://xxx.lanl.gov/abs/arXiv:1508.07166}{{\tt
  arXiv:1508.07166}}.

\bibitem{Seo:2015yqp}
H.~Seo, ``{Status of RENO-50},'' {\em PoS} {\bf NEUTEL2015} (2015) 083.

\bibitem{Huber:2002rs}
P.~Huber, M.~Lindner, and W.~Winter, ``{Synergies between the first generation
  JHF-SK and NuMI superbeam experiments},'' {\em Nucl. Phys.} {\bf B654} (2003)
  3--29, \href{http://xxx.lanl.gov/abs/arXiv:hep-ph/0211300}{{\tt
  arXiv:hep-ph/0211300}}.

\bibitem{Winter:2013ema}
W.~Winter, ``{Neutrino mass hierarchy determination with IceCube-PINGU},'' {\em
  Phys.~Rev.} {\bf D88} (2013) 013013,
  \href{http://xxx.lanl.gov/abs/arXiv:1305.5539}{{\tt arXiv:1305.5539}}.

\bibitem{Blennow:2013vta}
M.~Blennow and T.~Schwetz, ``{Determination of the neutrino mass ordering by
  combining PINGU and Daya Bay II},'' {\em JHEP} {\bf 09} (2013) 089,
  \href{http://xxx.lanl.gov/abs/arXiv:1306.3988}{{\tt arXiv:1306.3988}}.

\bibitem{Feruglio:2002af}
F.~Feruglio, A.~Strumia, and F.~Vissani, ``{Neutrino oscillations and signals
  in $\beta$ and $0\nu2\beta$ experiments},'' {\em Nucl.~Phys.} {\bf B637}
  (2002) 345--377, \href{http://xxx.lanl.gov/abs/arXiv:hep-ph/0201291}{{\tt
  arXiv:hep-ph/0201291}}.

\bibitem{Fong:2013wr}
C.~S. Fong, E.~Nardi, and A.~Riotto, ``{Leptogenesis in the universe},'' {\em
  Adv.~High Energy Phys.} {\bf 2012} (2012) 158303,
  \href{http://xxx.lanl.gov/abs/arXiv:1301.3062}{{\tt arXiv:1301.3062}}.

\bibitem{MSW-W}
L.~Wolfenstein, ``Neutrino oscillations in matter,'' {\em Phys.~Rev.} {\bf D17}
  (1978) 2369--2374.

\bibitem{MSW-MS}
S.~Mikheyev and A.~Y. Smirnov, ``Resonant neutrino oscillations in matter,''
  {\em Prog.~Part.~Nucl.~Phys.} {\bf 23} (1989) 41--136.

\bibitem{Akhmedov}
E.~K. Akhmedov, A.~Dighe, P.~Lipari, and A.~Y. Smirnov, ``Atmospheric neutrinos
  at {Super-Kamiokande} and parametric resonance in neutrino oscillations,''
  {\em Nucl.~Phys.} {\bf B542} (1999) 3--30,
  \href{http://xxx.lanl.gov/abs/arXiv:hep-ph/9808270}{{\tt
  arXiv:hep-ph/9808270}}.

\bibitem{Parke:2015goa}
S.~Parke and M.~Ross-Lonergan, ``{Unitarity and the three flavour neutrino
  mixing matrix},'' \href{http://xxx.lanl.gov/abs/arXiv:1508.05095}{{\tt
  arXiv:1508.05095}}.

\bibitem{Agafonova:2015jxn}
{\bf OPERA} Collaboration, N.~Agafonova {\em et~al.}, ``{Discovery of $\tau$
  Neutrino Appearance in the CNGS neutrino beam with the OPERA experiment},''
  {\em Phys. Rev. Lett.} {\bf 115} (2015) 121802,
  \href{http://xxx.lanl.gov/abs/arXiv:1507.01417}{{\tt arXiv:1507.01417}}.

\bibitem{Rott:2015kwa}
C.~Rott, A.~Taketa, and D.~Bose, ``{Spectrometry of the Earth using Neutrino
  Oscillations},'' \href{http://xxx.lanl.gov/abs/arXiv:1502.04930}{{\tt
  arXiv:1502.04930}}.

\bibitem{Winter:2015zwx}
W.~Winter, ``{Atmospheric neutrino oscillations for Earth tomography},'' {\em
  Nucl.\ Phys.\ B} {\bf 908} (2016) 250,
  \href{http://xxx.lanl.gov/abs/1511.05154}{{\tt 1511.05154}}.

\bibitem{Abbasi:2011ss}
{\bf IceCube} Collaboration, R.~Abbasi {\em et~al.}, ``{IceCube Sensitivity for
  Low-Energy Neutrinos from Nearby Supernovae},'' {\em Astron. Astrophys.} {\bf
  535} (2011) A109, \href{http://xxx.lanl.gov/abs/arXiv:1108.0171}{{\tt
  arXiv:1108.0171}}. [Erratum: Astron. Astrophys.563,C1(2014)].

\bibitem{Demiroers:2011am}
M.~Salathe, M.~Ribordy, and L.~Demirors, ``{Novel technique for supernova
  detection with IceCube},'' {\em Astropart. Phys.} {\bf 35} (2012) 485--494,
  \href{http://xxx.lanl.gov/abs/arXiv:1106.1937}{{\tt arXiv:1106.1937}}.

\bibitem{Bruijn:2013ibl}
R.~Bruijn, ``{Supernova Detection in IceCube: Status and Future},'' {\em Nucl.
  Phys. Proc. Suppl.} {\bf 237-238} (2013) 94--97,
  \href{http://xxx.lanl.gov/abs/arXiv:1302.2040}{{\tt arXiv:1302.2040}}.

\end{thebibliography}\endgroup
